\documentclass[reprint,secnumarabic,amssymb,nobibnotes,aip,jcp,superscriptaddress]{revtex4-1}

\usepackage{bm}
\usepackage{color}
\usepackage{graphicx}  
\usepackage{inputenc}
\begin{document}

\title{Thermoelectric energy recovery at ionic-liquid/electrode interface}

\author{Marco Bonetti}
\author{Sawako Nakamae}
\author{Bo Tao Huang}
\affiliation{Service de Physique de l'Etat Condens\'e, CEA-IRAMIS-SPEC, 
CNRS-UMR 3680, CEA Saclay, F-91191 Gif-sur-Yvette Cedex, France}
\author{Thomas J.~Salez}
\affiliation{\'Ecole des Ponts ParisTech, 6 et 8 avenue Blaise Pascal, 
Champs-sur-Marne F-77455 Marne-la-Vall\'ee, France}
\affiliation{Service de Physique de l'Etat Condens\'e, CEA-IRAMIS-SPEC, 
CNRS-UMR 3680, CEA Saclay, F-91191 Gif-sur-Yvette Cedex, France}
\author{C\'ecile~Wiertel-Gasquet}
\affiliation{Service de Physique de l'Etat Condens\'e, CEA-IRAMIS-SPEC, 
CNRS-UMR 3680, CEA Saclay, F-91191 Gif-sur-Yvette Cedex, France}
\author{Michel~Roger}
\affiliation{Service de Physique de l'Etat Condens\'e, CEA-IRAMIS-SPEC, 
CNRS-UMR 3680, CEA Saclay, F-91191 Gif-sur-Yvette Cedex, France}

\date{(\today)}

\begin{abstract}
A  \textcolor{black}{Thermally Chargeable Capacitor}
 containing a binary solution of 1-ethyl-3-methylimidazolium 
bis(trifluoromethylsulfonyl)-imide (EMIMTFSI) in acetonitrile is 
electrically charged by 
applying a temperature gradient to two ideally polarisable electrodes. The
corresponding thermoelectric coefficient is -1.7~mV/K for 
platinum foil electrodes and -0.3~mV/K for nanoporous carbon electrodes. 
Stored electrical energy is extracted by discharging the capacitor 
through a resistor. 
\textcolor{black}{The measured capacitance of the electrode/ionic-liquid 
interface is 5~$\mu$F
for each platinum electrode while it becomes four orders of magnitude 
larger $\approx$~36~mF for a single nanoporous carbon electrode.}
Reproducibility of the effect through repeated
 charging-discharging cycles under a 
steady-state temperature gradient demonstrates the robustness of the 
electrical charging process at the liquid/electrode interface. The 
acceleration of the charging by convective flows 
is also observed. This offers the possibility to convert
waste-heat into electric energy without exchanging electrons between ions 
and electrodes, in contrast to what occurs in  most thermogalvanic cells. 

\end{abstract}

\maketitle

\section{Introduction}
Conversion of low-grade waste heat (typically at temperatures $< 200^o$ C) into 
electrical energy has become a topical challenge in energy research of 
the 21st century.
For example, solid state  thermoelectric converters are attracting 
attention as promising devices for  portable
generators.\cite{Kuroki} An alternative to solid-state thermoelectric
 generators can be found in thermogalvanic cells containing an electrolyte and 
a redox couple\cite{Burrows,Richter,Hornut,Kuzminskii,Hinoue,Gunawan,Yamato,Abraham,Pringle,Zinovyeva,Salazar,Uhl,Quickenden,Yang-2014} where the
 difference in the redox potentials (reaction entropies) at two electrodes 
under a thermal gradient can be used to generate electricity. 
 In thermogalvanic cells, ions circulate inside
 the cell and exchange electrons with the electrodes.
 Presently, high thermogalvanic coefficient values are found
in ionic-liquid/organic solvent mixtures\cite{Galinski} using various 
inorganic and organic redox 
couples.\cite{Abraham,Pringle,Zinovyeva}
The redox reaction can also take place inside  electrodes via
insertion/deintercalation of ions,  \emph{e.g.} lithium in an
 intercalated compound like Li$_x$V$_2$O$_5$, leading to a thermally
 rechargeable battery.\cite{Amatucci}

It has recently  been proven by Qiao's 
group\cite{Qiao,Xu,Lim-2012,Lim-2013} that the thermocells 
can also be conceived with pure electrolytes (without 
redox couples) that operate with \emph{ideally polarizable} electrodes
with no exchange of electrons between electrodes and the electrolyte ions.
In their experiments, platinum and 
nanoporous carbon electrodes were immersed in a cell made of two compartments
maintained at different temperatures and 
electrically connected through a salt bridge.
In this case, no electrical current flows inside the cell and thus the 
thermocell operates by a
pure capacitive process at the electrodes. 
\textcolor{black}{Therefore the cell operates as
a \emph{``Thermally Chargeable Capacitor''} ({TCC}).}
The thermoelectric potential is then due to the 
internal electric field\cite{Agar} generated by the thermal drift of ions 
(Soret effect) in the electrolyte and the temperature dependence 
of the surface potential at the liquid/electrode interfaces. At the contact
with electrodes, ions lose a part of their solvation layers and their
 entropy is modified. The corresponding thermoelectric potentials 
for aqueous solutions of alcaline salts have been compared in
 Ref. \onlinecite{Lim-2012}.
The electrical energy stored in the form of accumulated charge at the 
electrode interfaces can be subsequently recovered by
'discharging' through an external load.

In this paper, we extend the work of Qiao's group\cite{Lim-2012,Lim-2013} to 
the interface between \emph{ideally polarisable} electrodes and an ionic liquid.
 Ionic liquids (IL) have a low melting temperature close to room temperature 
unlike most molten salts, a relatively high decomposition 
temperature\cite{Galinski,Ngo} and a large electrochemical 
window.\cite{Hayyan} They can operate on a wide temperature range 0--200$^oC$,
 suitable for low-grade waste heat recovery applications. 

The IL/metal electrodes interface is known to be complex, extending over 4--8 
layers, giving rise to a vast literature 
(for a short review on this subject see \emph{e.g.}
 Ref.~\onlinecite{Su}). IL's large ions can be adsorbed on a metal surface 
in multiple configurations.\cite{Mao}
Such conformational changes are accompanied by a partial loss of entropy 
from the bulk state, possibly 
resulting in a large temperature-dependent surface potential. Here,
we have chosen the binary mixture 1-ethyl-3-methylimidazolium 
 bis(trifluoromethylsulfonyl)-imide (EMIMTFSI, 391.3 g/mole),
diluted to 2M in 
acetonitrile (AN). \textcolor{black}{This concentration gives a weak maximum of 
the Seebeck coefficient and the highest ionic 
conductivity $\sigma$(40$^o$C)= $(57.3 \pm 0.5)$~mS~cm$^{-1}$ among 
different organic solvents.\cite{McEwen} }

To estimate the electrical energy stored in the \textcolor{black}{TCC} 
under a temperature gradient, we have performed
discharge potential measurements by connecting the electrodes to an
 external resistive load. Bare platinum foil electrodes 
were first used as a model system which has been extensively studied both
experimentally and theoretically.
 Repeated cycles of charge/discharge under a constant
 temperature gradient show that the charging of the IL/electrode 
interface layer is a robust process.
 The same experiments were then repeated with platinum-foil
electrodes covered with a thin sheet of nanoporous carbon, 
which increases the electrode capacitance by four orders of magnitude, 
a necessary condition for heat to electrical energy conversion applications.

\section{Experimental setup}
\begin{figure}[!tp]
\begin{center}
\includegraphics[width=0.45\textwidth]{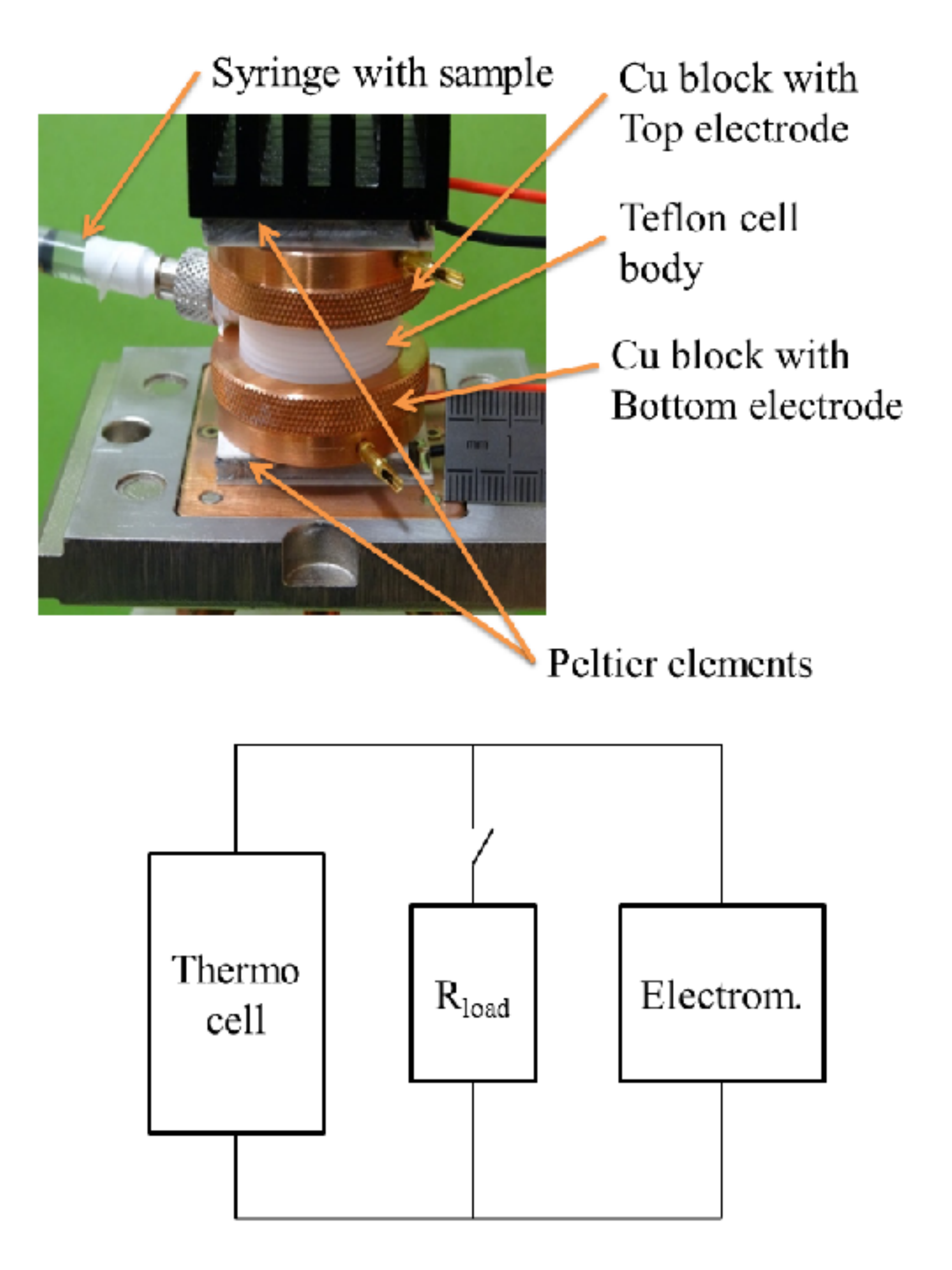}
\caption{Top:\textcolor{black}{ Thermally Chargeable Capacitor}.
 Two copper blocks  screwed onto
 the Teflon cell body 
press the platinum  foil electrodes against two O-rings to achieve sealing.
 The copper blocks are electrically in contact with the electrodes and are 
thermally regulated by means of Peltier elements. The electrodes are 
connected to an electrometer for open-circuit potential measurement, or 
to a resistive load via a switch-box to discharge the cell. The cell is 
25 mm in height and 35 mm in diameter. Notice the syringe that is used to fill 
the cell and allows the fluid to expand when the cell is heated.
Bottom: corresponding electrical circuit (see text for explanations).}
\label{photo}
\end{center}	
\end{figure}
The experimental setup (\textcolor{black}{TCC} geometry) is different from 
that of Lim \textit{et al}.\cite{Lim-2012} Rather than two horizontally
 connected compartments held at different temperatures, here the 
differential potential measurements were carried out in a 
 single-volume vertical cell with a $\phi=6$~mm diameter sample cavity 
machined out from a solid Teflon 
cylinder closed by two symmetrical 10 mm diameter platinum foil electrodes of
 100 $\mu$m thickness (99.99 \% pure, Sigma-Aldrich). (A complete view of 
the cell assembly is shown in Figure~1.).
\textcolor{black}{Our geometry is thus similar to that corresponding to
 usual supercapacitors.}
 The Pt electrode surface is wet polished with a diamond paste with
 particle sizes reduced from 9 to 1~$\mu$m, subsequently cleaned in an 
ultrasonic bath with acetone or acetonitrile and rinsed in ultrapure water and
 then vacuum dried. Additional cleaning by immersion of the electrodes
 in a concentrated HCl (36\%wt) solution
 for many hours and ultrasonication in ultrapure water
 did not have any effect on the measured thermoelectric potential. 
The electrode surface in contact with 
the liquid is $A \approx 0.28$~cm$^2$ and 
the distance between the electrodes is $l=6~$mm.  Two 10~mm thick copper 
blocks are screwed onto the cell 
body, thereby squeezing the Pt electrodes hermetically against the Teflon 
cylinder (the sample volume is 0.17 cm$^3$). To investigate the influence of 
the effective electrode surface, a thin sheet of nanoporous carbon 
(NanoTuneX Y-Carbon) with $\approx$20~nm average pore diameter was sandwitched
between the platinum foil and the cell body. 
The electrodes are electrically and thermally in contact with 
the top and bottom copper blocks which ensures a fast heat transfer to and 
from the liquid mixture. The temperature of the Cu blocks is monitored by 
100~$\Omega$-platinum resistance thermometers and regulated by a temperature
 controller (LakeShore 340) that finely adjusts the output power of  two 
  Peltier elements with lapped alumina surface
 of 30x30~mm$^2$. The thermal stability
 of both Cu blocks, $\delta$T, is less than 2~mK over many hours.
The natural convection of the liquid mixture can be either minimized; 
by heating only the top Cu-block while keeping the bottom Cu-block at 30~$^o$C, 
or maximized; by reversing the heating direction.   
The maximum allowed temperature difference is  $\approx 60^o$C, 
corresponding to a temperature 
gradient across the cell body $\approx 65$~K~cm$^{-1}$. 
The sample liquid filling 
is achieved by injection  through a small radial hole in the cylindrical 
cavity using a syringe. The syringe is left in place 
so that overfilling of the cell when the two Cu blocks are screwed together
or the thermal expansion of the fluid during the measurements 
 is compensated by the displacement of 
the piston inside the syringe. The electric conductivity of 
the sample was also measured using the same cell.  
The ionic liquid (EMIMTFSI, 98\% purity, Sigma-Aldrich) and acetonitrile
 (99.8\% purity, Sigma-Aldrich) were used as received without further
 purification. The \textcolor{black}{TCC} is filled with a solution of 2M EMIMTFSI in AN.
The water content was less than 0.05\% w/w (Karl-Fisher).

The entire cell is electrically floating with respect to the Peltier elements
 and is placed in a Faraday cage. The thermoelectric open-circuit potential 
(during charging) and the discharge potential between the electrodes are 
measured by a high impedance electrometer with a 
$10^{14}~\Omega$ input resistance (Keithley K-6514). Discharging is achieved 
via a manual switch that connects instantaneously 
the cell electrodes to an external resistive load  that ranges 
between 1.5~k$\Omega$ to 100~M$\Omega$. 
The electrical circuit corresponding to the measurement setup is presented
in Figure 1 (bottom panel).

\section{Experimental results with platinum-foil electrodes}

\subsection{Thermoelectric coefficient}

The open circuit potential  $\Delta V$ between the hot and cold electrodes 
(no current flow) is proportional
to the temperature difference $\Delta T$ between the two electrodes:
\begin{equation}
\Delta V = - \Lambda\Delta T.
\end{equation} 
$\Lambda$ represents the \emph{thermoelectric coefficient}.
$\Delta V$ can be split into three contributions:
\begin{equation}
\Delta V=  \phi_H - E_{int}l-\phi_C
\end{equation}
where $\phi_H$ and  $\phi_C$ represent the interface electrostatic potentials
at the hot and cold electrodes, $l$ is the electrode separation
 and $ E_{int}$ is a uniform Seebeck internal 
electric field resulting from the thermal drift of ions: 
\begin{equation}
\mathbf{E}_{int}= S_e\mathbf{}\nabla T.
\end{equation}
$S_e$ represents the Seebeck coefficient.
$\phi_H$ and $\phi_C$ are the results of IL anions and  cations adsorbed
 at the metal interfaces. Such adsorption effects have been observed by 
scanning tunelling microscopy,\cite{Pan,Mao} 
atomic force microscopy,\cite{Wakeham} angle-resolved X-ray photoelectron 
spectroscopy,\cite{Cremer} electrochemical impedance 
spectroscopy.\cite{Silva,Lockett}
The temperature dependence of the structure of an 
IL/metal interface has also been
evidenced in Ref.~\onlinecite{Wakeham,Silva,Lockett}, which implies that the
interface potential $\phi$ depends on temperature as well. For
small temperature variations, we can write:
\begin{equation}
\phi_H-\phi_C=\frac{d\phi}{dT}\Delta T
\end{equation}
and we have:
\begin{equation}
\Lambda=S_e-\frac{d\phi}{dT}
\label{lambda}
\end{equation}
We expect the second term
in Eq.~(\ref{lambda}) to be particularly important in the case of large-ion
adsorption since $d\phi/dT$ depends on
$\Delta S_{conf}/e$, where  $\Delta S_{conf}$ is the change
in configurational entropy of adsorbed ions.\cite{Agar,Bonetti}

For monovalent ions, the Seebeck coefficient of an electrolyte at 
initial state, \emph{i.e.} when a 
temperature gradient is applied while the ion concentration is still
uniform throughout the cell, is:\cite{Agar} 
\begin{equation}
S_e^{init}=(t_+\widehat{S}^+-t_-\widehat{S}^-)/e
\end{equation}
where $\widehat{S}^+$, $\widehat{S}^-$ are the Eastman entropies of transfer
of cations and anions and $t_+,t_-$ the corresponding Hittorf transport
numbers. When ions have moved to establish a Soret steady-state,
in an open circuit condition, we have:\cite{Agar,Wurger}
\begin{equation}
S_e^{st}=(\widehat{S}^+-\widehat{S}^-)/(2e)
\end{equation}

\begin{figure}[!t]
\begin{center}
\includegraphics[width=0.45\textwidth]{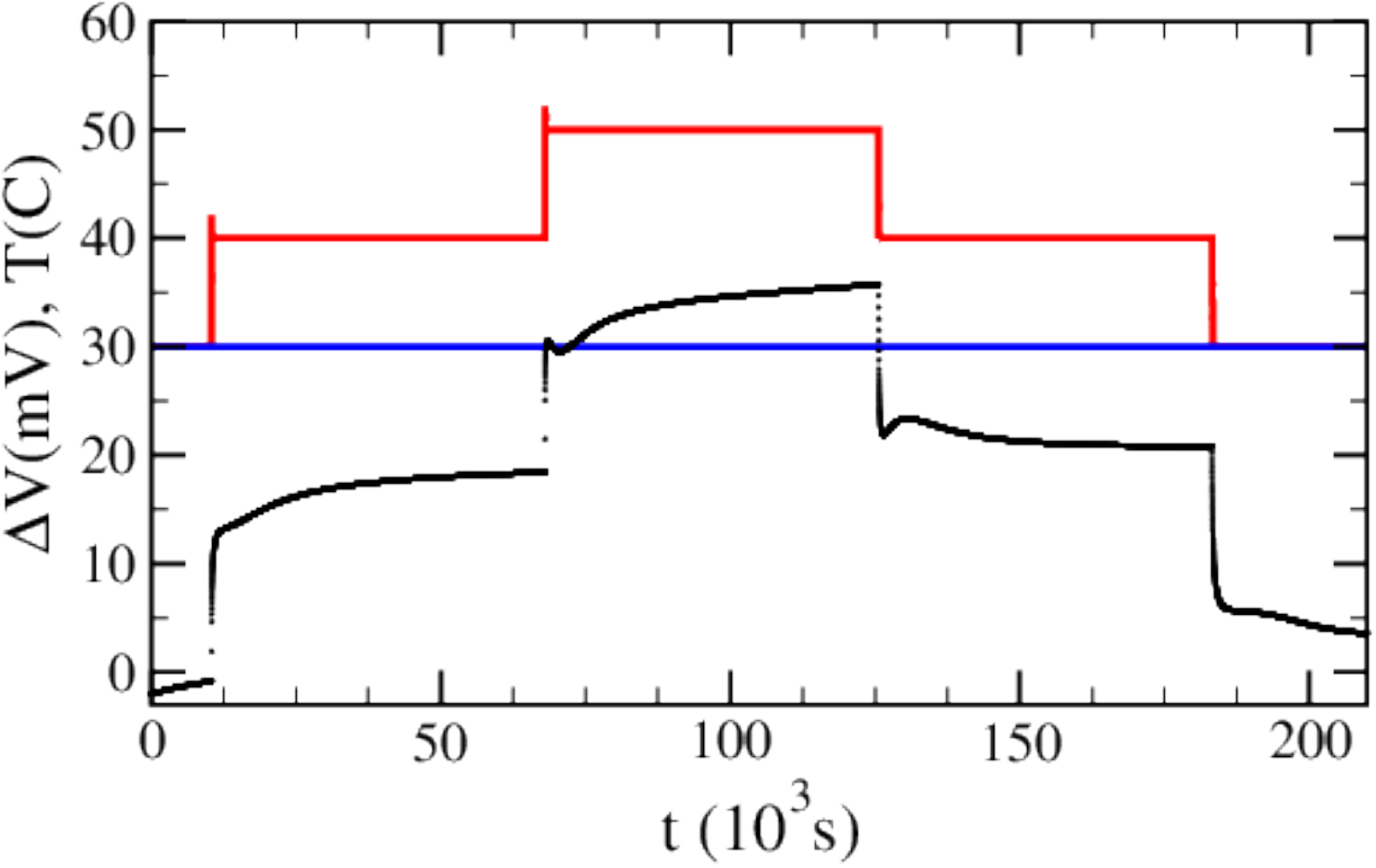}
\caption{Thermoelectric potential $\Delta V$ (black line) for a solution of 
EMIMTFSI (2M) in acetonitrile when the temperature of the top electrode
 (red line) is changed by a 10~$^o$C step between 30$^o$C and 50$^o$C. The
temperature of the bottom electrode (blue line) is maintained at 30$^o$C. }
\label{reversibility}
\end{center}	
\end{figure}

\begin{figure}[!b]
\begin{center}
\includegraphics[width=0.4\textwidth]{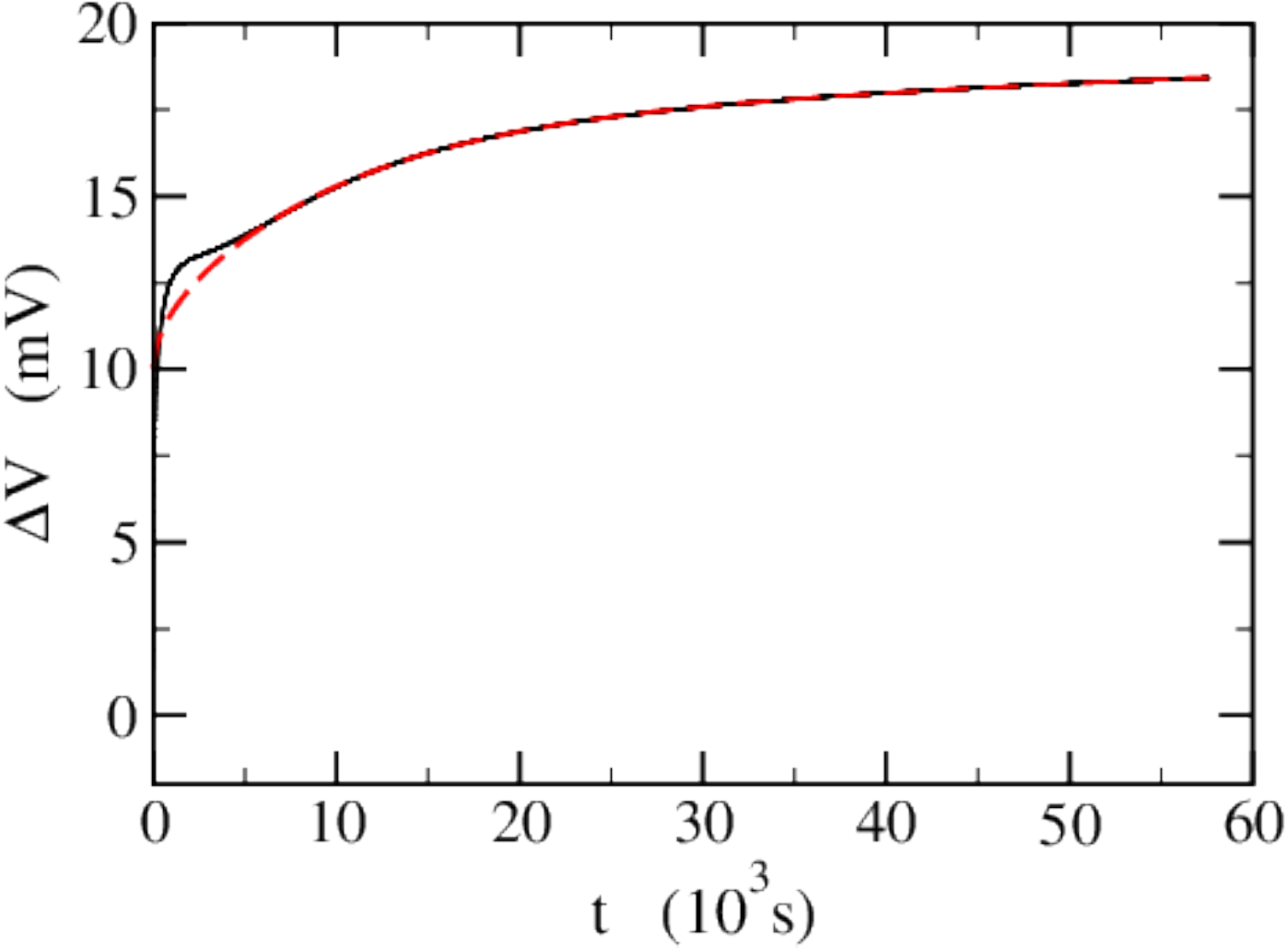}
\caption{Zoom on the first step in Figure~\ref{reversibility}
 (solid black line).
The red dashed line is a  fit through Eq.~(\ref{time-equation}). }
\label{time-evolution}
\end{center}	
\end{figure}

A typical time evolution of $\Delta V$ when the temperature of the hot 
electrode is heated from 30$^o$C to 50$^o$C by 10$^o$C steps, while the cold 
electrode
is maintained at 30$^o$C is shown in Figure~\ref{reversibility}. A nice
reversibility is observed when cooling the hot electrode back to 30$^o$C. An 
average value $\Lambda_{Pt}\approx -1.7$~mV~K$^{-1}$ 
is deduced from the measurement.
The time evolution of $\Delta V$ corresponding to the first heating cycle is 
zoomed in Figure~\ref{time-evolution}.
This evolution is 
governed by the Soret diffusion of IL ions through the cell which, in 
one dimension, leads to:\cite{Bierlein,Turner}
\begin{eqnarray}
\Delta V(t)-\Delta V(0)=& (\Delta V(\infty)-\Delta V(0))\times \nonumber \\
&\left[1-\frac{8}{\pi^2}\sum_{n~\rm{odd}}\frac{\exp (-n^2t/\theta)}{n^2}\right]
\label{time-equation}
\end{eqnarray}
where $\theta=l^2/(\pi^2D)$ is the characteristic time of diffusion with $D$ 
representing the diffusion coefficient of cations. 
For $t>\theta$ only the lowest order term in the sum of exponential terms is 
significant.

The experimental data at $t > 10000~$s (Figure~\ref{time-evolution}) have 
been fit to Eq.~(\ref{time-equation}). 
An additional linear term corresponding to an instrumental drift 
of 74 $\mu$V/hour was also taken in account. 
The characteristic time $\theta$ is $\approx 10620~$s$~\approx 3$~hours from 
which we estimate $D\approx 3.4\times10^{-10}~$m$^2$~s$^{-1}$. 
This value is three times higher than the value deduced from nuclear magnetic 
resonance measurements in pure EMIMTFSI at room temperature.\cite{Noda}  
Such an increase in $D$ is in agreement with the
 numerical simulations\cite{Chaban}
 that predict a 4-fold boost in the diffusion coefficient of 2M
EMIMBF4 diluted in acetonitrile.
In Figure~\ref{time-evolution}, the red dashed line represents the 
extrapolation of the \textit{full} Eq.~(\ref{time-equation}) down to $t=0$. 
Note that the time required for the onset of a uniform temperature gradient
 is 5 to 10~mn, which explains the deviation from 
Eq.~(\ref{time-equation}) at $t< 1000~$s.

\subsection{\textcolor{black}{TCC} charging and discharging}

Discharging of the \textcolor{black}{TCC} is obtained by connecting the two electrodes 
to an external resistive load (R$\approx$~6--100~M$\Omega$) much larger than 
the internal resistance of the liquid sample (37~$\Omega$ at 40$^o$C). 
The experimental procedure for charging/discharging measurements is described 
as follows: 

\begin{figure}[!b]
\begin{center}
\includegraphics[width=0.4\textwidth]{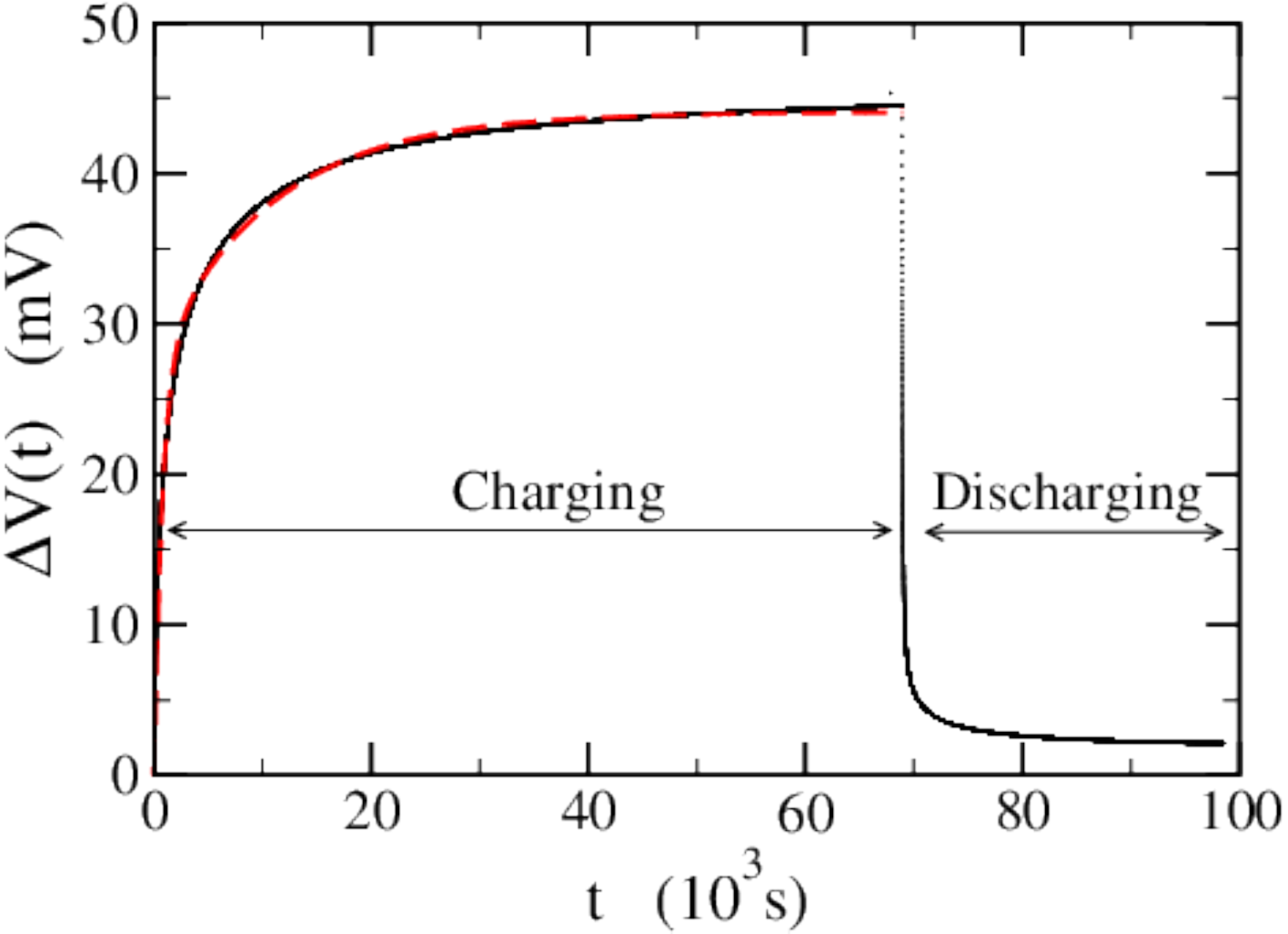}
\caption{Charge/discharge potential of the \textcolor{black}{TCC} with 
Pt foil electrodes as a function of time.
At time $t=0$, the \textcolor{black}{TCC} is completely 
discharged and charging potential is 
measured in open circuit (cell connected to the electrometer alone)
during 19~hours up to a steady-state value. Discharged is then
accomplished by connecting the electrodes to a 20 M$\Omega$ resistor.
During the whole process the top and bottom electrode are maintained 
at  50$^o$C and 30$^o$C respectively. The dashed red line is 
 fit to a sum of two exponentials with characteristic times
$\alpha=790~$s and $\theta =10600~$s.
}
\label{charge-discharge}
\end{center}	
\end{figure}

\begin{enumerate}
\item
Initial discharge: The cell is completely discharged through a low value 
resistor.
\item
Charging phase: At an experimentally defined $t=0$, the cell is put in open
 circuit (disconnected from the resistive load).
From $t=0$ to $t\approx19~$hours we observe an increase of the open-circuit
 potential $\Delta V$ from 0 to a steady value of 
$\approx$~45~mV (see Figure~\ref{charge-discharge}, charging phase).
At this stage, the \textcolor{black}{TCC} is considered to be fully charged. 
$\Delta V(t)$ can be fit by the sum of two exponential terms with
 characteristic times $\alpha = 790~\rm{s}\approx 13~$min and
$\theta = 10600~\rm{s}\approx 3~$hours. $\theta$ is close to the
 time-scale corresponding to the ion diffusion through the cell (see previous
subsection).
\item
Discharging phase: The electrodes are then connected to an external resistive 
load (20~M$\Omega$ for the measurement shown in Figure~\ref{charge-discharge}). 
The potential $\Delta V$ drops rapidly to a low value $\approx 2$~mV. 

\end{enumerate}
Note that a steady-state temperature gradient between the top (50$^o$C) and
 the bottom (30$^o$C) electrode is maintained throughout the measurements.

\begin{figure}[!b]
\begin{center}
\includegraphics[width=0.4\textwidth]{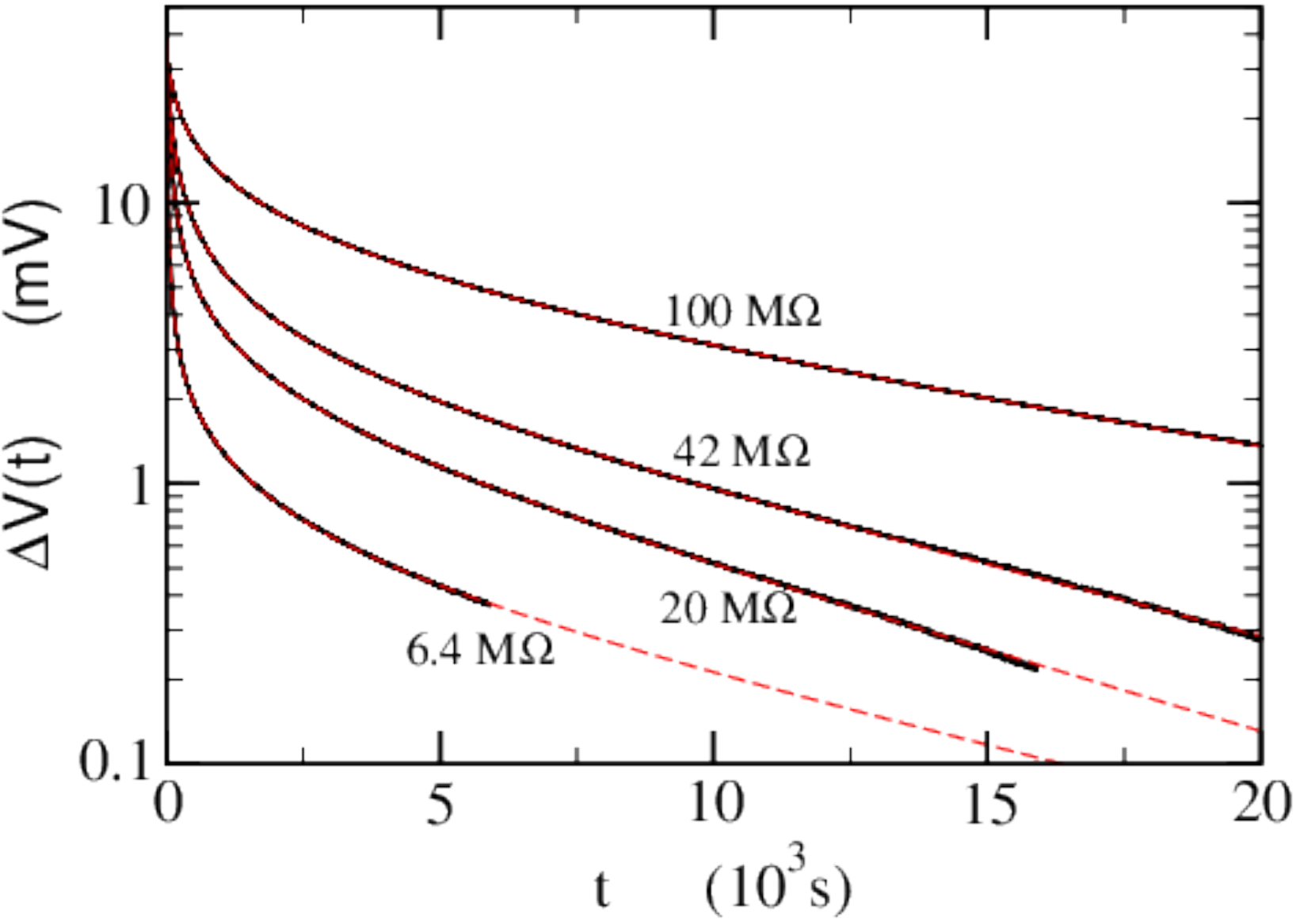}
\caption{Discharge potential through different loads from 6.4 to 100~M$\Omega$. 
Solid black lines: experimental data. Red dashed lines: 
fits using Eq.~(\ref{pade}).
\textcolor{black}{TCC} with Pt-foil electrodes, filled with EMIMTFSI (2M) in acetonitrile.
A small instrumental offset at large $t$ has been subtracted from 
the experimental data.}
\label{discharges}
\end{center}	
\end{figure}

 Similar charge/discharge curves are obtained with the resistive loads
 $R$ = 6.4, 42 and 100~M$\Omega$. The discharge potentials are reported in 
a semi-log plot in Figure~\ref{discharges}. As can be seen from the figure, 
the semi-logarithmic plots of  $\ln \Delta V(t)$ are not linear, thus
 a rational fraction containing two time constants was used to fit the data. 
 The time constants  $\tau$ and $\theta$ characterize respectively 
the fast (discharging) and slow (thermodiffusion) processes:
\begin{equation}
-\frac{t+at^2+bt^3}{\tau+ct+\theta bt^2} \label{pade}
\end{equation}

The long time-scale constant $\theta\approx 3\pm 1~$hours is roughly
 independent of the resistive load. 
It is comparable to the diffusion time of ions through the cell and may be
 due to the readjustment of the ion concentration 
gradient in the bulk to satisfy the new steady-state boundary conditions
 corresponding to 
equipotential electrodes rather than open-circuit potential electrodes.

\begin{figure}[!t]
\begin{center}
\includegraphics[width=0.4\textwidth]{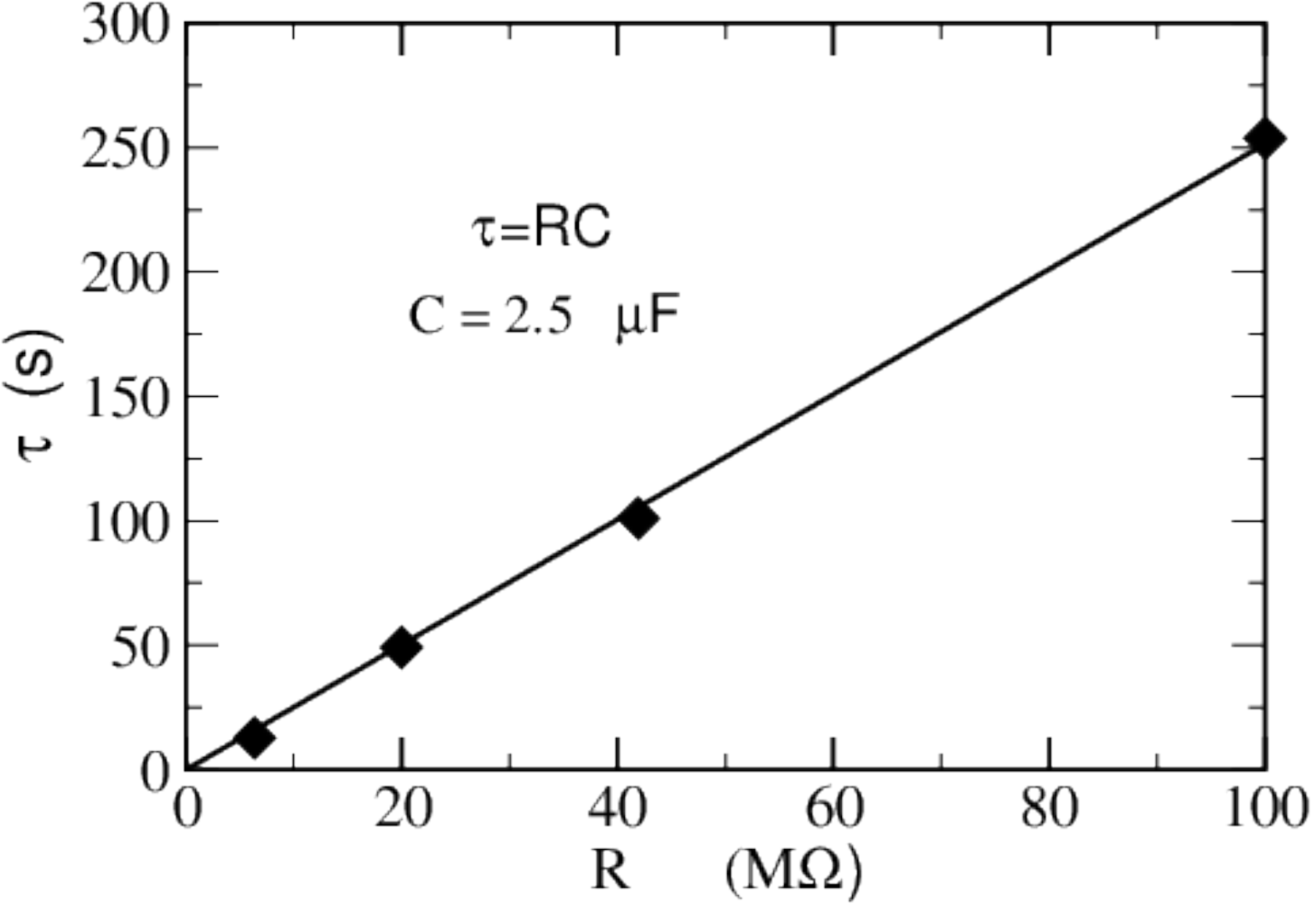}
\caption{Short relaxation time $\tau$ corresponding to different
 resistive loads $R$ = 6.4, 20, 42 and
 100~M$\Omega$.}
\label{taufig}
\end{center}	
\end{figure}

The corresponding values of the short relaxation time $\tau$ are reported as 
a function of the load resistance $R$ in Figure~\ref{taufig}. 
As seen from the graph, $\tau$ is proportional to $R$, from which one can
 extract capacitance $C=\tau/R\approx 2.5~\mu$F.\textcolor{black}{Assuming
 identical electrodes,
this corresponds to the capacitance of two electrode/ionic-liquid interfaces in 
series (see section IV). The capacitance of a single electrode is thus
$C_{Pt}=5~\mu$F, giving per surface area $C_{Pt}/A=18\mu\textrm{F/cm}^2$.}
This value is of the same order of magnitude as the differential capacitance 
obtained through electrochemical impedance spectroscopy
at the interface of  similar ILs with noble-metal electrodes, 
\emph{e.g.} $\approx 4~\mu\rm{F}cm^{-2}$ with 
1-buthyl-3-methylimidazolium hexafluorophosphate
(BMIMPF6) on platinum electrodes,\cite{Silva} and $\approx 6-8~\mu\rm{F}cm^{-2}$
with 1-ethyl-3-methylimidazolium tris(pentafluoroethyl)-trifluorophosphate
(EMIMFAP) on gold electrodes.\cite{Roling} 
This indicates that the short relaxation time $\tau$ is indeed related to the 
fast rearrangement of the ions at the liquid/solid interface, while
 negligible current flows through the bulk sample.

\begin{figure}[!b]
\begin{center}
\includegraphics[width=0.4\textwidth]{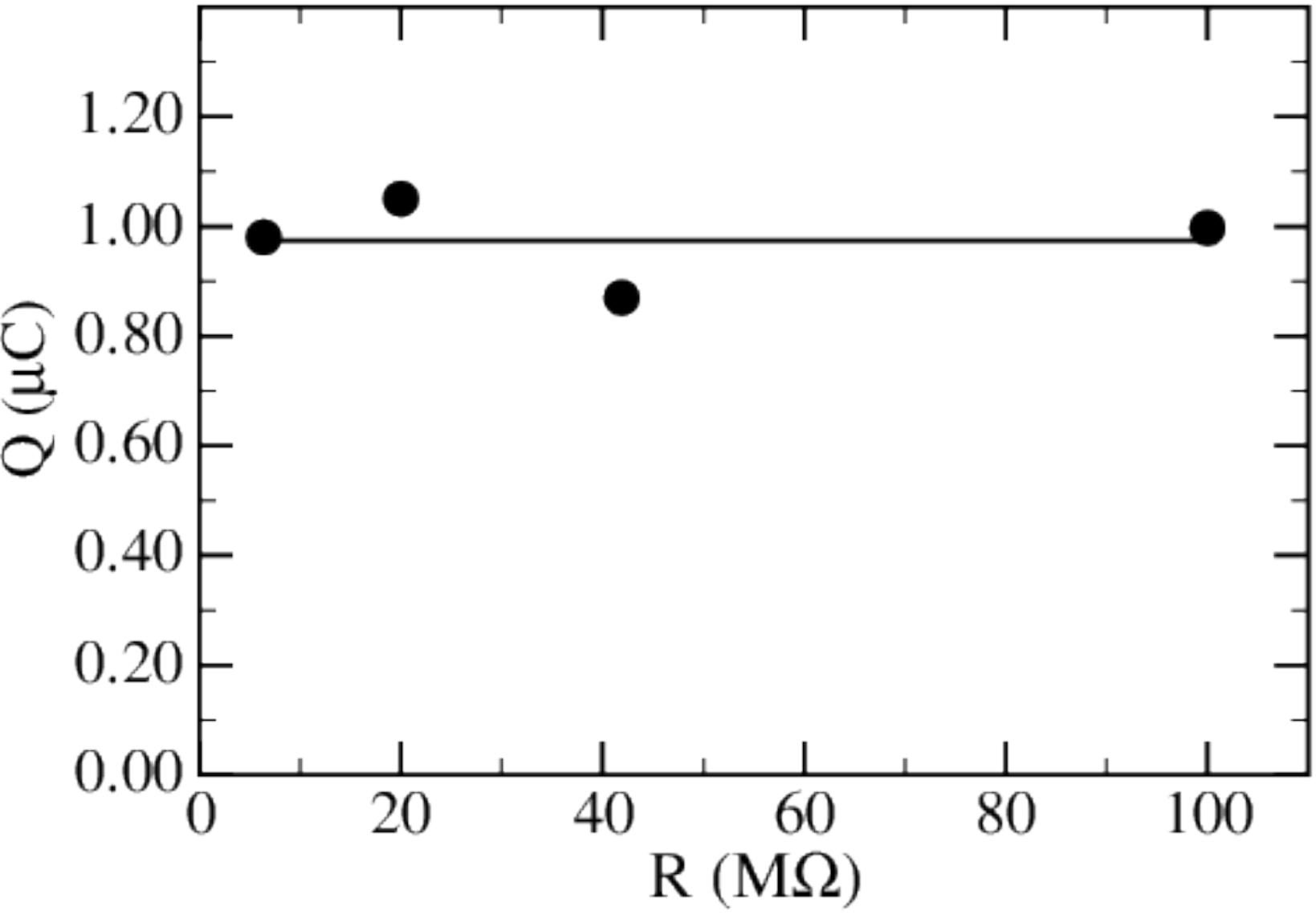}
\caption{Charge Q accumulated in the IL/Pt-electrode interface as a 
function of the resistive load R.}
\label{charge}
\end{center}	
\end{figure}

From the discharge potential the  current flowing through the
 resistive load can be calculated. The time integration of the electric
 current gives an estimate on the total charge $Q$ accumulated in the 
electrodes. Figure~\ref{charge} shows an almost constant value 
of the charge $Q\approx 1~\mu$C for all $R$, corresponding to a charge 
density ($Q/A$) of $\approx 3~\mu$C~cm$^{-2}$.

\subsection{Charge/Discharge cycling at constant temperature gradient}

\begin{figure}[!b]
\begin{center}
\includegraphics[width=0.4\textwidth]{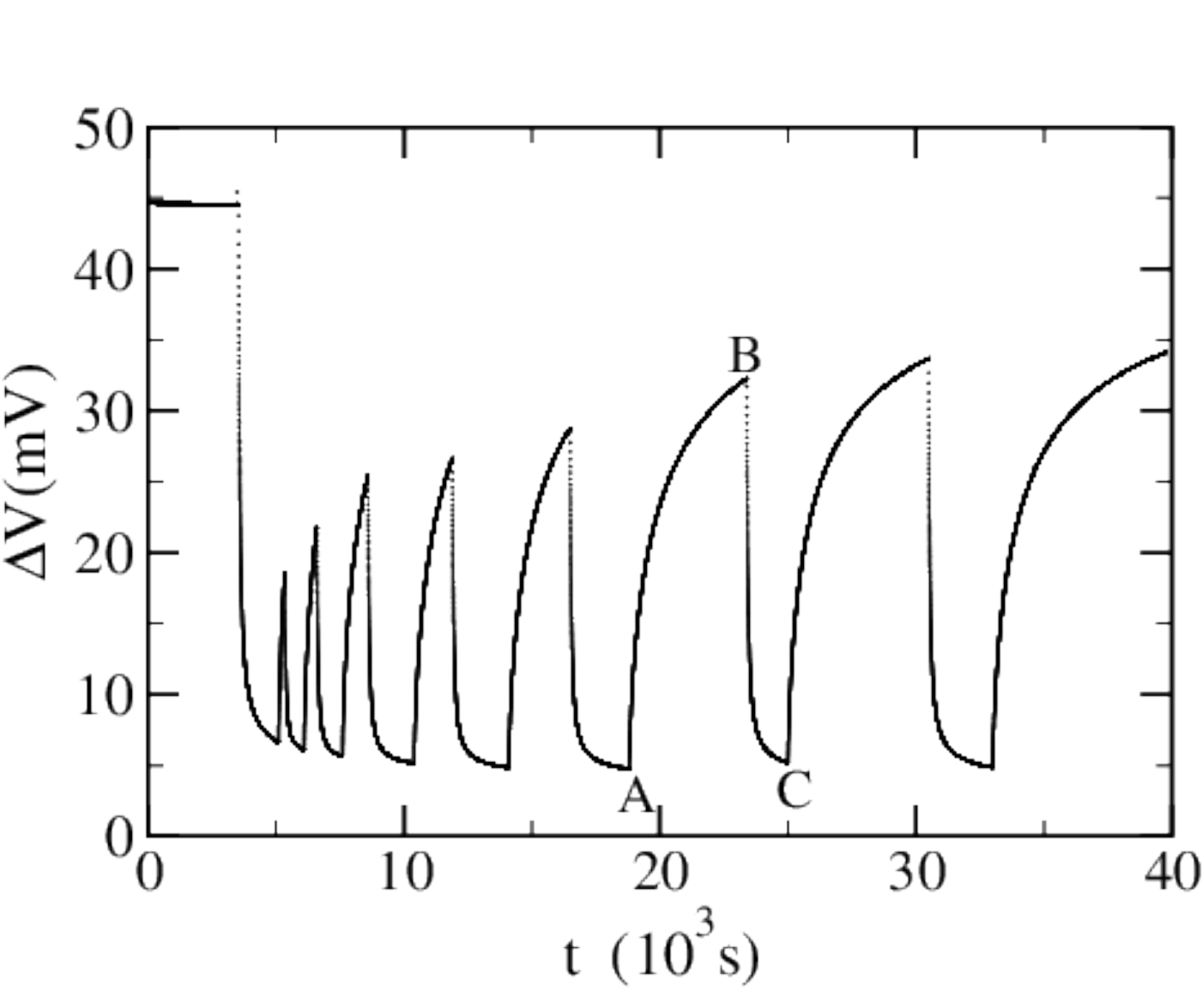}
\caption{Charging and discharging cycles of the \textcolor{black}{TCC} with Pt-foil
electrodes, filled with 2~M
EMIMTFSI in acetonitrile.
Charging: curve AB. Discharging: curve BC. The time duration $(t_B-t_A)$ 
of charging is 
progressively increased. Discharging is made over a 20M$\Omega$ resistive load. 
Steady-state temperature difference is 20K. Heating is from the top 
electrode; heat flow is purely diffusive.}
\label{cycles}
\end{center}	
\end{figure}

\begin{figure}[!t]
\begin{center}
\includegraphics[width=0.4\textwidth]{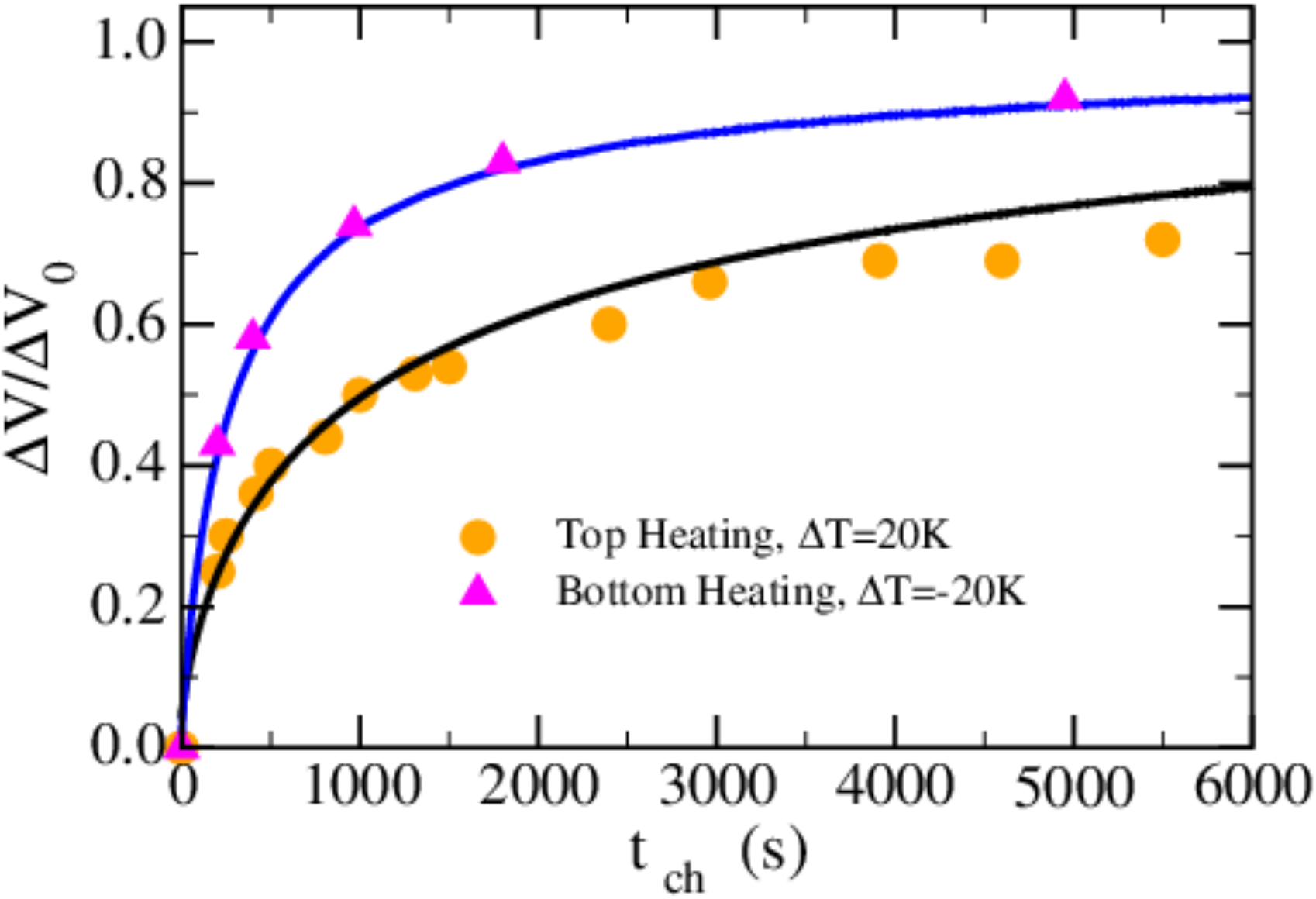}
\caption{Normalized charging potential $\Delta V/\Delta V_0$ as a function 
of  time $t_{ch}$. Filled circles: Heating on the top electrode; pure diffusive
 heat flow. Filled triangles: Heating from the bottom electrode; convective 
heat flow. In both experiments, the steady-state temperature difference
 is 20~K. Full lines: Continuously time-recorded charging potential starting
 from a completely discharged cell and ending to the full charged state.}
\label{cycles-bilan}
\end{center}	
\end{figure}

To test the recharging capability of the \textcolor{black}{TCC},
 we have performed many charge/discharge  cycles  
with progressively increasing charging time $t_{ch}$ under 
a steady-state temperature gradient (see Figure~\ref{cycles}). 
A 20 M$\Omega$ resistive load was used for discharging. 
 Here, we have intentionally stopped charging/discharging before reaching the
stationary states. 
 Figure~\ref{cycles-bilan} shows the normalized differential potential
 $\Delta V/\Delta V_0$ as a function 
of $t_{ch}$, where $\Delta V_0$ is the stationary differential potential
 measured when the electric layer is fully charged. 
 Two sets of data are represented. The circles are obtained for pure
 diffusive heat flow
 when the temperature of the top electrode is heated to 50$^o$C and the 
bottom is maintained close to room temperature (30$^o$C). 
The triangles are  obtained when an internal convective flow  settles in the
 cell heated from the bottom with the same temperature difference. 

When heated from the bottom, the onset of convective  flow occurs at a 
critical Rayleigh number 
$Ra_c\approx 1700$.\cite{Drazin}
For pure EMIMTFSI using the current cell geometry and $\Delta T$ = 20~K 
between the electrodes, $Ra\approx 20\times 10^3$ well above the critical value.
In our less viscous binary solution, this value is expected to be even higher.
Consequently, heat transfer across the bottom and top electrodes is dominated 
by an almost turbulent convective heat flow rather than pure heat
 conduction through a stationary liquid.\cite{Drazin} 
It can be observed from Figure~\ref{cycles-bilan} that 50\% of the fully 
charged state is achieved in 
$\Delta t\approx 250$~s when there are convective flows, whereas it 
takes $\approx 1000$~s when the heat flow is purely diffusive. Clearly,
 convective flows accelerate the charging of the electric layer. 
In Figure~\ref{cycles-bilan} we have also plotted (full solid lines)
 the continuous differential voltage measurement during the (full) charging
 phase after completely discharging the \textcolor{black}{TCC}. 
Both curves (with and without convection) coincide remarkably well with the
 measurements performed in the charge/discharge sequence,
showing the robustness of the electric layer charging process that can 
withstand many cycles. 

\subsection{Efficiency}

The electrical energy stored at the IL/electrode interface and recovered
 through a
single discharge is: $W=(1/2)C\Delta V^2=(1/2)C\Lambda_{Pt}^2 \Delta T^2$. 
With $\Lambda_{Pt}\approx -1.7$~mV/K at
 $\Delta T=20^o$C, we obtain $W/A\approx 0.05~\rm{mJ~m}^{-2}$. 
During a charging time of $\Delta t\approx 250$~s (see section III.3.),
 the energy loss  $\mathcal{Q}$
is essentially due to the heat flow from the hot to
the cold electrode: $\mathcal{Q}\approx \kappa A\Delta t\Delta T/l$, where
 $\kappa \approx 0.3~\rm{W~m}^{-1}~\rm{K}^{-1}$ is the thermal conductivity of
 the electrolyte and $l=6$~mm
the distance between electrodes. 
The \textcolor{black}{TCC} efficiency is expressed by:
\begin{equation}
\eta=\frac{W}{\mathcal{Q}}=\frac{Tl\Lambda_{Pt}^2}{2\kappa\Delta t}\frac{C}{A}\eta_c
\end{equation}
where $\eta_c=\Delta T/T$ represents the Carnot efficiency. We obtain
$\eta\approx 0.24\times 10^{-9}~\eta_c$, which is far too small for any
 imaginable waste-heat recovery application. 
However, the efficiency can be improved by adjusting various experimental
 parameters:
\begin{enumerate}
\item
The electrode surface: Electrodes with larger effective surface area 
can be used to increase the specific capacity.
\item Thermoelectric coefficients: $\Lambda$ of the order of
 10~mV~K$^{-1}$
 has been observed in solutions of tetraalkylammonium nitrate 
in alcohols.\cite{Bonetti}
 Since the efficiency is proportional to $\Lambda^2$ a factor 100 might be
gained by prospecting among the huge number of combinations IL/solvent.
\item The (re)charging time $\Delta t$ can be shorter depending
 on the IL and the nature of the electrodes.
\item \textcolor{black}{TCC} geometry: The thermal heat loss may be substantially reduced
by modifying the geometry of the cell (\emph{e.g.} non uniform section with
reduced diameter at mid distance between electrodes, 
or by intercalating a porous membrane between electrodes). 
\end{enumerate}

To illustrate, we have investigated one such example; namely
 the charging/discharging effect using nanoporous carbon electrodes,
 as presented in the following section.

\section{Experimental results with nanoporous-carbon electrodes}

\textcolor{black}{A  0.2~mm thick and 4~mg weight nanoporous carbon 
(NanoTuneX Y-Carbon) electrode foil has been sandwitched between the
platinum electrode and the cell body at both ends. N$_2$ adsorption/desorption 
isotherms were performed with a Micromeritics ASAP 2020 instrument. BET
analysis gives a surface area of $\approx 300~\textrm{m}^2/g$. The pore-size
distribution is determined from desorption isotherm according to the BJH
method. The average pore size is $\approx 20\pm 10~\textrm{nm}$.}

The thermoelectric coefficient under the same condition as in the previous
 section is $\Lambda_C =-0.3~$mV~K$^{-1}$,
approximately $6$ times smaller than that obtained with Pt electrodes: 
$\Lambda_{Pt} =-1.7~$mV~K$^{-1}$. 
This observation shows the importance of the interface potential $\phi(T)$ 
in Eq.~(\ref{lambda}) and its
drastic dependence on the nature of the electrode. 
 The remarkably large thermoelectic
coefficient $\Lambda$ on platinum electrodes is likely due to 
 the configurational entropy change in EMIM$^+$ and/or TFSI$^-$ ions 
adsorbed at the surface. 

A series of charging/discharging cycles through various resistive loads
from 1.5 to 10~k$\Omega$ has been performed. Note that, due to much larger 
capacitances of carbon electrodes, it was necessary to use much
smaller R values here than those used with Pt electrodes.

\begin{figure}[!t]
\begin{center}
\includegraphics[width=0.4\textwidth]{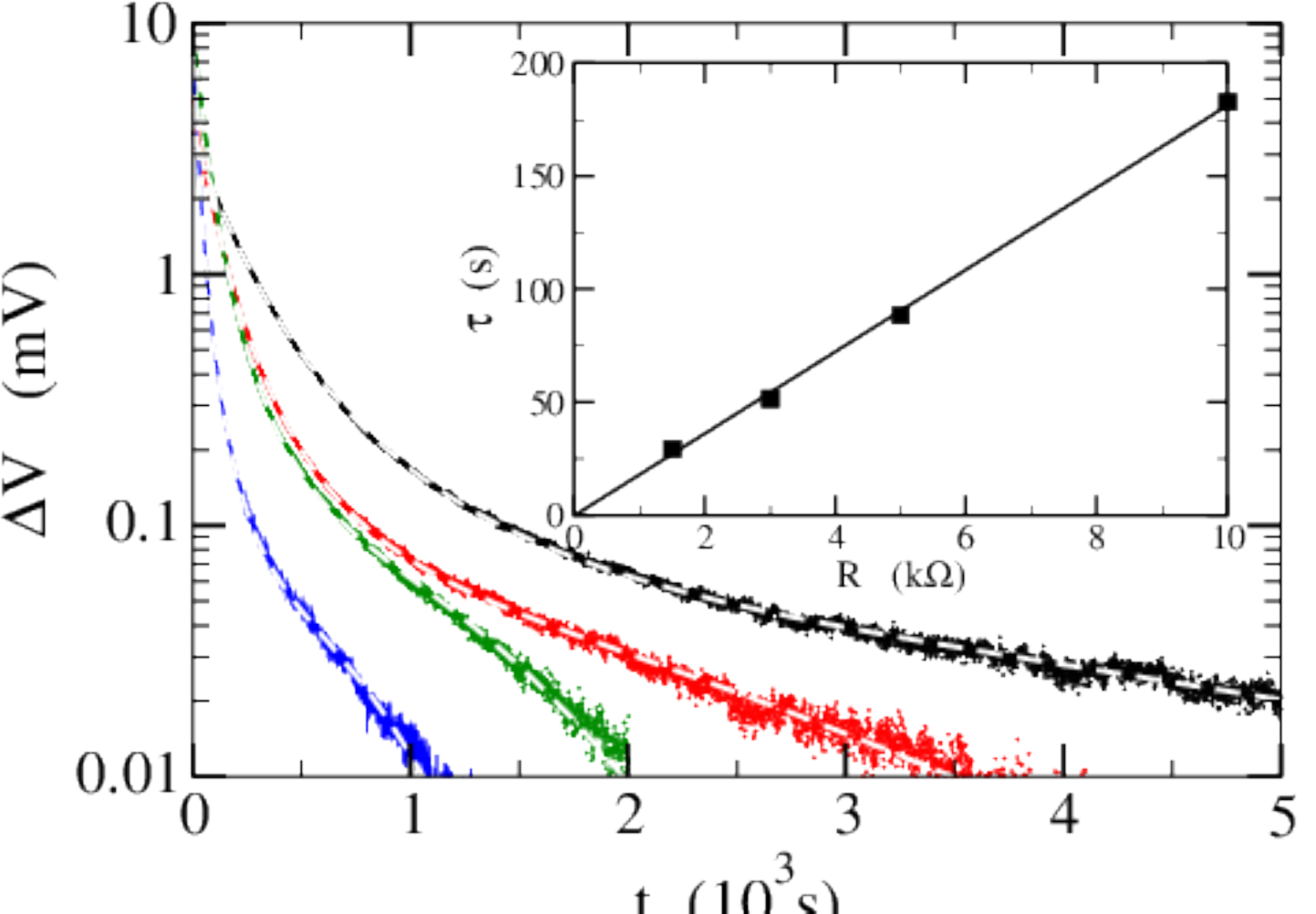}
\caption{\textcolor{black}{TCC} with Y-Carbon electrodes. Discharge through different loads:
 1.5 (blue), 3 (green), 5 (red), and 10~k$\Omega$ (black). The white dashed
lines are fits through Eq.~(\ref{pade}). The corresponding
short relaxation times $\tau$ as a function of $R$ are
show in inset. The corresponding capacitance is 18.3 mF, four orders of 
magnitude larger than that obtained with Pt electrodes. Sample is 2M
EMIMTFSI in acetonitrile.}
\label{Ycarbonfit}
\end{center}	
\end{figure}

The discharge potentials are shown in Figure~\ref{Ycarbonfit}. 
The results are fit once again to Eq.~(\ref{pade}) 
to extract $\tau$ and $\theta$. 
The corresponding short relaxation-times $\tau$ are reported in the inset, from 
which the differential capacitance of the system was determined as $C=18.3$~mF, 
four orders of magnitude larger than that corresponding to Pt electrodes. 
The long time-scale constant $\theta\approx 2$~hours was obtained from
 $R=10$~k$\Omega$ measurement; (note that for smaller $R$, the experimental
time range ($t<\theta$) does not permit accurate determination of $\theta$).

\begin{figure}[!t]
\begin{center}
\includegraphics[width=0.4\textwidth]{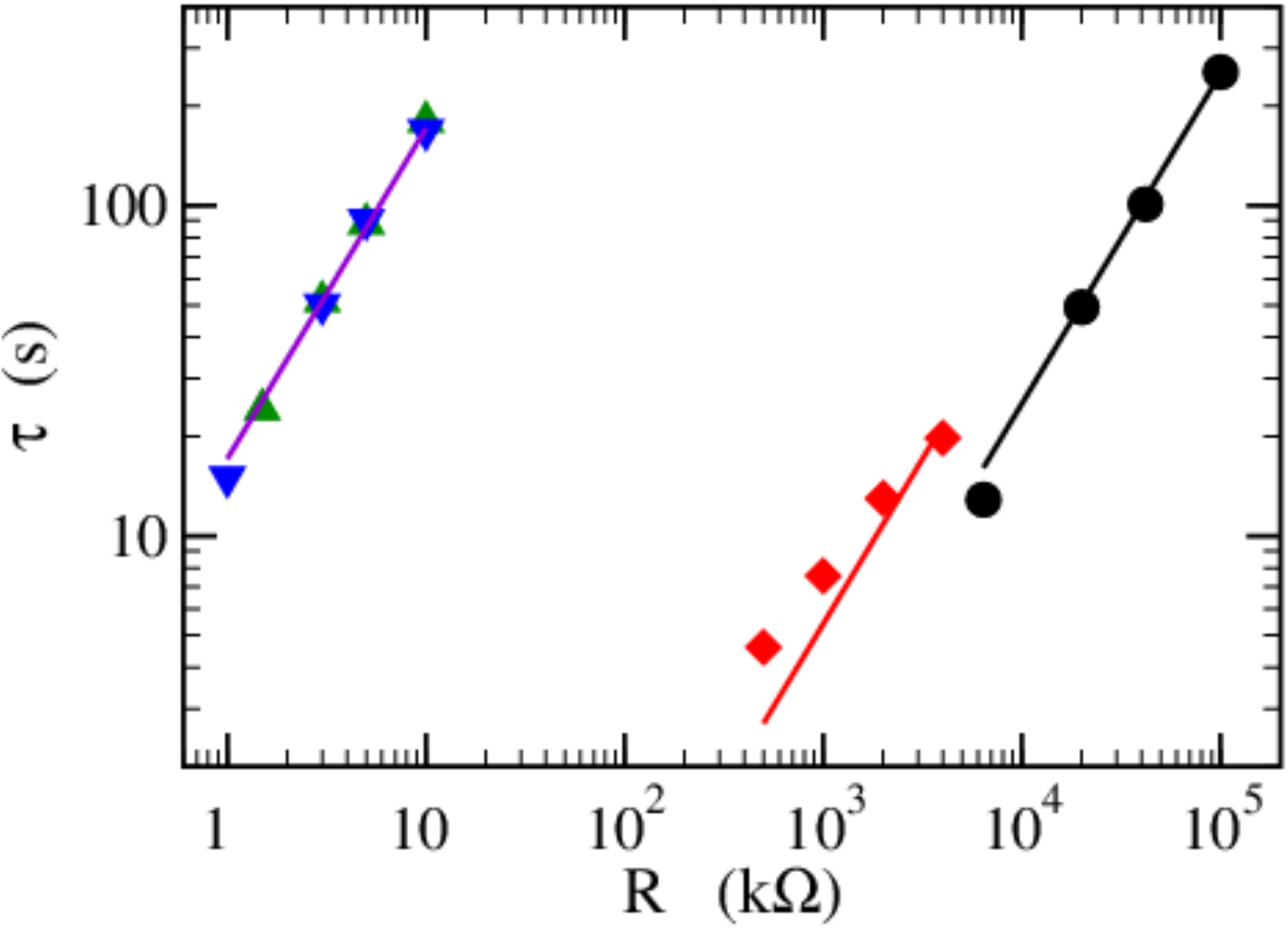}
\caption{Short charateristic times $\tau$ with: a) Two Pt electrodes (black
circles); the black line corresponds to two capacitors 
with $C_{Pt}=5~\mu$F in series;
b) Two Y-Carbon nanoporous electrodes at 30--50$^o$C (green up-triangles)
and  30--70$^o$C (blue down-triangles); the blue line corresponds to two
capacitors with $C_C=36.5~$mF in series; c) One Pt electrode and  one
Y-Carbon nanoporous electrode (red diamonds); the red line corresponds to
two different capacitances  $C_{Pt}=5~\mu$F  and  $C_C=36.5~$mF  in series.}
\label{lastfig}
\end{center}	
\end{figure}

Applying a larger temperature difference $\Delta T=40^o$C with $T_{top}=70^o$C
and  $T_{bot}=30^o$C leads  to the same conclusion.
In Figure~\ref{lastfig}, the short characteristic times $\tau$
obtained using the two different $\Delta T$ values are compared.
 They lead to the same capacitance $C=18.3$~mF.
Therefore, by simply changing the electrode material, we were able to increase 
the \textcolor{black}{TCC} efficiency by  $2 \approx 3$ orders of magnitude.

Finally, we have performed a set of charge/discharge with asymmetric electrodes:
\emph{i.e.}, nanoporous Y-Carbon electrode at the top, 
and Pt electrode at the bottom. The results are similar to 
those obtained with two symmetric pure platinum
 electrodes with a slightly higher capacitance value of 5.4~$\mu$F.
The short characteristic times obtained under different measurement conditions
are summarised in  Figure~\ref{lastfig}. In all cases the cell can be modelled
as two capacitances in series, taking $C_C=36.5~$mF for a carbon electrode
and $C_{Pt}=5~\mu$F for a platinum electrode. 

\textcolor{black}{The specific capacitance for one carbon electrodes is
 $\approx 3\mu \textrm{F/cm}^2$, a value close to that deduced 
from Fig. 4 in Ref. \onlinecite{Barbieri}:  $\approx 4\mu \textrm{F/cm}^2$ from
electrochemical analysis of a 1~M solution of the ionic liquid 
tetraethylammonium tetrafluoroboride in acetonitrile.}

\section{Conclusion}
We have investigated the thermal-to-electric energy conversion possibility 
in \textcolor{black}{TCC}'s through charge accumulation
 in the electrical layers forming 
at the IL-electrode interface. 
The accumulated electric charge can be dissipated through an external load
in a discontinuous way through a charge/discharge sequence while maintaining 
a steady-state temperature gradient across the cell. Most of the discharge
occurs in a short time corresponding to a fast rearrangement of the 
IL/electrode interface. Repeated cycles of charge/discharge of
 the \textcolor{black}{TCC}
show that the charging process is robust and well reproducible.

With no ionic electric current flowing through the cell, the described energy
recovery process differs from that of most thermogalvanic cells.
Some advantages of the present method include avoiding Joule heating and the
reliance on the ionic conductivity, which tend to limit the efficiency of 
thermogalvanic cells.

Although the present efficiency value is not usable for any waste-heat recovery
application, it can be improved greatly by a careful selection of electrode 
materials, ionic liquid and cell designs.
\textcolor{black}{Indeed, from Fig. 4 in Ref. \onlinecite{Barbieri},  carbon
nanopowder with higher surface area of order 2000 m$^2$/g increases the  capacity 
by a factor 10 with respect to the presently used Y-Carbon. 
Other nanostructured electrodes, 
as nanoporous metals or carbon nanotubes can also be investigated.}
The charging time of the electric layer can also be reduced 
greatly by inducing convective flows. Last but not the least a plethora of
IL/electrode combinations can be explored to increase the thermoelectric
coefficient $\Lambda_e$ to maximize the voltage difference, and ultimately,
to increase the efficiency of thermoelectric capacitors.

\section*{Acknowledgments}

\textcolor{black}{We thank  D. Duet for technical support and M. Bombled
 from Laboratoire
L\'eon Brillouin (CEA-CNRS, Saclay) for performing the BET surface area and
 BJH pore-size measurements on nanoporous carbon.} This project 
is supported by Agence Nationale pour la Recherche  (France)  ANR-TEFLIC- 
Progelec  12-PRGE-0011-01.

%

\end{document}